\documentclass{article}
\usepackage{amsmath}
\usepackage{amssymb}
\usepackage{amsfonts}
\usepackage{bm}
\usepackage[dvips]{graphicx}
\newcommand{\figurewidth}{.65\linewidth} % for the draft

\newcommand{\rev}[1]{{#1}}

\begin{document}

%------------------------------------------------------------------------------
% title and abstract

\title{\bf Linear Viscoelasticity of Dumbbells Interacting 
via Gaussian Soft-Core Potential}
\author{Takashi Uneyama
\\
\\
JST-PRESTO, and 
Center for Computational Science, \\
Graduate School of Engineering, 
Nagoya University, \\
Furo-cho, Chikusa, Nagoya 464-8603, Japan}

\date{}

\maketitle

\begin{abstract}
{%
In polymer melts, the interaction between segments are considered to be
screened and the ideal Gaussian chain statistics is recovered. The experimental
fact that linear viscoelasticity of unentangled polymers can be well
described by the Rouse model is naively considered as due to this screening
effect. Although various theoretical models are based on the screening effect and the screening effect
is believed to be reasonable,
the screening effect cannot be fully justified on a solid theoretical basis.
In this work, we study the screening effect by utilizing a simple dumbbell type model.
We perform simulations for dumbbell systems in which particles interact via the Gaussian
soft-core potential. We show that, if the density of dumbbells is high,
the Gaussian soft-core interaction is actually screened and the static
structures are well described by the ideal model without Gaussian soft-core
interaction. We also show that the relaxation moduli of interacting
dumbbell systems approximately coincide to those of the non-interacting
dumbbell systems. In the low density systems, we observe the deviations
from the ideal non-interacting systems. For example, the relaxation
moduli become relatively broad. However, the relaxation moduli of such 
systems can be decomposed into the relaxation modes by the Gaussian soft-core
interaction and the bond. The bond relaxation mode can be successfully described
by a single Maxwell relaxation with effective relaxation strength and
time. Our results support a naive use of the Rouse model to analyze unentangled
polymer melts. \\
%\textbf{Key Words:} Linear viscoelasticity / Dumbbell model / Gaussian soft-core potential / Screening / Brownian dynamics
}
\end{abstract}

%------------------------------------------------------------------------------
% main text

%------------------------------------------------------------------------------
\section{INTRODUCTION}

Most of rheological properties of polymer melts can be attributed
to the dynamics of polymer chains. A polymer melt consists of many polymer
chains, and a polymer chain consists of monomers or segments. The segments
in a polymer melt interact each other via the interaction potential
such as the Lennard-Jones potential. Therefore, ultimately, the dynamics of a
polymer melt can be described as the dynamics of strongly interacting particles.
This description is formally correct but does not give us clear information on
the dynamics and rheological properties of polymer melts.
From this viewpoint, the dynamics of polymer chains will be too complex to
theoretically analyze. To theoretically analyze the dynamics of polymer chains
and compare some theoretical predictions with experimental data, we need to
introduce some approximations (or simplifications).

If we consider just a single tagged polymer chain in a melt, and ignore 
the interactions between segments (except the bonds), we can construct theoretical models
which are analytically tractable. The thus constructed models are often
called single-chain type models or mean-field models.
Naively, the ignorance of the interactions between segments is justified as follows. In a polymer melt, one tagged segment is surrounded
by other segments. The segments are expected to be uniformly distributed around
the tagged segment.
Then the force acting on the tagged segment will be
effectively canceled, and the segment will behave as if it feels
no interaction potential.
Such a screening picture is found in many textbooks\cite{Doi-Edwards-book,Kawakatsu-book,deGennes-book}, and various
properties of polymer melts are explained on the basis of the ideal chain
statistics which is based on the screening picture.
The screening picture can be experimentally justified by measuring
several physical quantities for polymer melts. For example, the chain conformation
in a polymer melt can be measured by scattering experiments and is known to be well
described by the ideal chain model.
This agreement is, however, theoretically not fully clear. The ignorance of the
interaction between segments is not theoretically justified, but rather
just assumed (at least currently). 

The screening effect seems to work for dynamics. If we assume the
dynamics of a single tagged chain in a polymer melt with the screening effect,
the dynamics can be described by a single-chain model such as 
the Rouse model\cite{Rouse-1953,Doi-Edwards-book}.
In the Rouse model, the tagged chain
is modeled as a bead-spring chain and the dynamics of the bead is simply
described by the overdamped Langevin equation. Various dynamical quantities
including the relaxation modulus can be straightforwardly calculated for
the Rouse model. The relaxation modulus predicted by the Rouse model
agrees well with the experimental relaxation modulus of a unentangled polymer melt.
The Rouse type behavior is also observed in molecular dynamics simulations\cite{Likhtman-Sukumaran-Ramirez-2007}.
Thus, at least apparently, the screening picture seems to be reasonable.

However, the dynamics of strongly interacting particles is generally not
that simple, and the applicability of the screening effect to the dynamics
of polymer melts is not theoretically clear.
Let us consider a single tagged chain in an unentangled polymer melt.
This tagged polymer chain is interacting with surrounding polymer chains. When
the tagged chain moves, it should push the surrounding chains to make vacancy.
Then the dynamics of the tagged chain and the surrounding chains should be
correlated, and we may observe the chain dynamics which is qualitatively
different from the Rouse dynamics. The fact that the dynamics can be well
described by the Rouse model implies that the dynamic correlation is effectively
canceled in polymer melts. To the best of the authors' knowledge, no theory
successfully justifies this cancellation. The polymer chains in unentangled
melts are rather strongly interacting, and it would be extremely difficult
to study the dynamics analytically without approximations.
Even in the case of simple liquids which have no intramolecular conformational
degrees of freedom, the theoretical models become quite complicated\cite{Hansen-McDonald-book}.

In this work, we study some aspects of the screening effect by
simulations for simple systems. Sometimes realistic
polymer models are too detailed and complex to study physical mechanisms.
Thus we limit ourselves to a simple model, although it may not be realistic.
To express the relaxational dynamics of polymer chains, the dumbbell
type model would be a suitable minimal model. Despite its simplicity,
the dumbbell type model can describe various interesting dynamical
behaviors of polymers\cite{Kroger-2004}
For example, by incorporating the modulation of the mobility or the friction coefficient
under flow, various nonlinear rheological behavior can be reproduced analytically\cite{Uneyama-Horio-Watanabe-2011,Watanabe-Matsumiya-2017,Watanabe-Matsumiya-Sato-inprint}.
Even if we limit ourselves to the linear viscoelastic behavior, dumbbell models with
fluctuating diffusivity (or the fluctuating friction coefficient) \cite{Uneyama-Miyaguchi-Akimoto-2019,Miyaguchi-Uneyama-Akimoto-2019}
or with the inertia effect \cite{Uneyama-Nakai-Masubuchi-2019} exhibit non-trivial interesting behavior.
Therefore, we employ dumbbells inceracting via some potentials as a model system
to study the screening effect.
\rev{Since the dumbbell model is highly coarse-grained, the Lennard-Jones potential (or similar potentials with cores)
is not suitable. Besides, we want to study how the strength of the interaction potential affects the
screening effect. For our purpose, a soft coarse-grained interaction potential with tunable interaction strength is
required. In this work we employ the Gaussian soft-core interaction potential.
The Gaussian soft-core interaction allows polymers to fully overlap, and we can
easily tune the interaction strength.
}
We perform simulations for systems with interacting dumbbells
and study some static and dynamic properties including the linear viscoelasticity.
We show that the screening of the interaction
is actually observed in such a simple model. Especially, if the density of 
the system is sufficiently high, the static structures of a single
tagged dumbbell in the system can be well described by the non-interacting, ideal dumbbell model
without interaction. Also, the linear viscoelasticity of an interacting dumbbell system is well described by the
corresponding ideal dumbbell system.
We discuss how we can interpret
the simulation results, and also discuss the validity of the naive screening picture for other coarse-grained polymer models.

%------------------------------------------------------------------------------
\section{MODEL}
\label{model}

In this work, we employ the Brownian dynamics model originally developed for
symmetric block copolymers\cite{Uneyama-2009}. We divide a single polymer chain into two subchains
with the equal molecular weight. Then, we assume that a subchain is fully
equilibrated and segments are distributed around the center of mass of the
subchain. Then, the subchain would be interpreted as a soft particle with
a Gaussian type kernel\cite{Louis-Bolhuis-Hansen-Meijer-2000,Addison-Hansen-Krakoviack-Louis-2005}. A polymer chain can be modeled as a dumbbell type
molecule in which two soft particles are connected by a tethering potential.

We consider a system with the volume $V$ and number of polymer chains $M$.
The number density of the particles is calculated as $\rho = 2 M / V$.
We express the position of two soft particles in the $j$-th polymer
as $\bm{R}_{j,1}$ and $\bm{R}_{j,2}$, respectively.
The dimensionless units are convenient for simulations.
We take the average size of the bond vectors for polymers,
the thermal energy $k_{B} T$ ($k_{B}$ and $T$ are the Boltzmann constant and the temperature),
and the friction coefficient for a subchain, to be unity.
In the dimensionless units, the interaction potentials can be
expressed as:
\begin{equation}
 \label{interaction_potential}
 \begin{split}
 U[\lbrace \bm{R}_{j,k} \rbrace]  = & \sum_{j} \phi(\bm{R}_{j,1} - \bm{R}_{j,2}) \\
 &  + \sum_{(j,k) > (l,m)} v(\bm{R}_{j,k} - \bm{R}_{l,m}) ,   
 \end{split}
\end{equation}
\begin{align}
 \label{tethering_potential}
 \phi(\bm{r}) & = \frac{1}{2} \bm{r}^{2} , \\
 \label{gaussian_core_potential}
 v(\bm{r}) & = \frac{\epsilon}{\pi^{3/2}} \exp(-\bm{r}^{2}).
\end{align}
In the second summation in eq~\eqref{interaction_potential}, the expression $(j,k) > (l,m)$
represents that the summation is taken only for $2 j + k > 2 l + m$.
The potential functions $\phi(\bm{r})$ and $v(\bm{r})$ represent the 
tethering interaction and the Gaussian soft-core interaction, respectively.
$\epsilon \ge 0$ represents the strength of the Gaussian soft-core interaction.
We may call $\epsilon$ as the interaction strength in what follows.
(In the case of $\epsilon = 0$, there is no interaction between different
polymer chains and thus the system reduces to the ideal dumbbell model.
\rev{If the dumbbells do not interact each other, the system reduces to
the canonical ensemble of harmonic dumbbells.})

As the dynamics model, we employ a simple overdamped Langevin equation.
In absence of the external flow field, the dynamic equation is given
as the following Langevin equation (in dimensionless units):
\begin{equation}
 \label{langevin_equation}
 \frac{d\bm{R}_{j,k}(t)}{dt} = - \frac{\partial U(\lbrace \bm{R}_{j,k} \rbrace)}{\partial \bm{R}_{j,k}}
  + \sqrt{2} \bm{w}_{j,k}(t) .
\end{equation}
Here, $\bm{w}_{j,k}(t)$ is the Gaussian white noise which satisfies the
fluctuation-dissipation relation. The first and second order moments of
$\bm{w}_{j,k}(t)$ become
\begin{equation}
 \label{fluctuation_dissipation_relation_noise}
 \langle \bm{w}_{j,k}(t) \rangle = 0, \quad
 \langle \bm{w}_{j,k}(t) \bm{w}_{l,m}(t) \rangle = \delta_{jl} \delta_{km} \bm{1} \delta(t - t'),
\end{equation}
where $\langle \dots \rangle$ represents the statistical average and $\bm{1}$ is
the unit tensor. The numerical simulations are performed by discretizing the
Langevin equation \eqref{langevin_equation} and numerically integrate it.
We employ the second order
stochastic Runge-Kutta scheme\cite{Honeycutt-1992} to accurately integrate the Langevin equation.
The details of the numerical
scheme can be found in the previous work\cite{Uneyama-2009}.

The stress tensor of the system can be calculated as
\begin{equation}
 \label{stress_tensor}
  \begin{split}
   \hat{\bm{\sigma}} & = \frac{1}{V} \sum_{j,k,l,m} \frac{\partial U(\lbrace \bm{R}_{j,k} \rbrace)}{\partial \bm{R}_{j,k}}
  \bm{R}_{j,k} - \rho \bm{1} \\
   & = \frac{1}{V} \sum_{j}
   (\bm{R}_{j,1} - \bm{R}_{j,2}) (\bm{R}_{j,1} - \bm{R}_{j,2}) \\
   & \qquad + \frac{\epsilon}{V \pi^{3/2}} \sum_{(j,k) > (l,m)}
   \exp[- (\bm{R}_{j,k} - \bm{R}_{l,m})^{2}] \\ 
   & \qquad \times (\bm{R}_{j,k} - \bm{R}_{l,m}) (\bm{R}_{j,k} - \bm{R}_{l,m}) - \rho \bm{1}  .
  \end{split}
\end{equation}
If we ignore the contribution of the Gaussian soft-core interaction,
the stress tensor reduces to the widely utilized form which is consistent with the stress-optical rule.
In this work, we want to explore how the Gaussian soft-core interaction affects
the linear viscoelasticity, and thus we utilize eq~\eqref{stress_tensor}
without any further approximations.

In equilibrium, the stress tensor becomes isotropic on average. This can be
related to the pressure of the system, $P$, as
\begin{equation}
 \label{pressure_and_stress}
  \left\langle \hat{\bm{\sigma}} \right\rangle_{\text{eq}} = - P \bm{1} ,
\end{equation}
where $\langle \dots \rangle_{\text{eq}}$ represents the statistical average
over the equilibrium distribution. If the Gaussian soft-core interaction is
absence, the pressure can be exactly calculated because the system behaves as an ideal gas.
But for the interacting dumbbell systems,
we should introduce some approximations to obtain the analytic expression.

\rev{The linear viscoelasticity of this model can be numerically calculated
by the Green-Kubo formula.}
The stress tensor is a function of the particle positions and thus it can be
interpreted as a function of time.
The Green-Kubo formula states that
the shear relaxation modulus is given as the correlation function of
the stress tensor\cite{Evans-Morris-book}:
\begin{equation}
 \label{green_kubo_relation}
 G(t) = V \left\langle \hat{\sigma}_{xy}(t) \hat{\sigma}_{xy}(0) \right\rangle_{\text{eq}} .
\end{equation}
We numerically calculate the relaxation
modulus by using eq~\eqref{green_kubo_relation} with eq~\eqref{stress_tensor}.

In order to estimate the contribution of the soft-core interaction
to various properties of an interacting dumbbell system, we consider
a system without the tethering potential as a reference system.
We may call this reference model as the soft particle model.
The soft particle system consists of $M$ particles, and the $j$-th particle
position is expressed as $\bm{R}_{j}$. The particle density of the system is
$\rho = M / V$.
The interaction energy of the soft particle model is given as
\begin{equation}
 \label{interaction_potential_soft_particle}
 U[\lbrace \bm{R}_{j} \rbrace] 
 = \sum_{j > k} v(\bm{R}_{j} - \bm{R}_{k}) , 
\end{equation}
instead of eq~\eqref{interaction_potential}. Also, the stress tensor
of the soft particle model is given as
\begin{equation}
 \label{stress_tensor_soft_particle}
  \begin{split}
   \hat{\bm{\sigma}} 
   & =  \frac{\epsilon}{V \pi^{3/2}} \sum_{j > k}
   \exp[- (\bm{R}_{j} - \bm{R}_{k})^{2}] \\ 
   & \qquad \times (\bm{R}_{j} - \bm{R}_{k}) (\bm{R}_{j} - \bm{R}_{k}) 
  - \rho \bm{1}  .
  \end{split}
\end{equation}
instead of eq~\eqref{stress_tensor}. Dynamic equation \eqref{langevin_equation}
is essentially the same (only the subscript $jk$ is changed to $j$).

We perform simulations with several systems with different particle
density and the interaction strength. We fix the system size to be $8^{3}$
and apply the periodic boundary condition to all the directions. 
We change the number of polymers $M$
and the interaction parameter $\epsilon$. We vary $M$ so that the particle
density $\rho$ varies as $\rho = 0.25, 0.5, 1, 2, $ and $4$.
We also
vary $\epsilon$ as $\epsilon = 1, 2, 4, 8, 16, 32,$ and $64$. For comparison,
we perform simulations for non-interacting systems by simply setting $\epsilon = 0$.
Also, for comparison, we perform simulations for soft particle systems which
has the same parameters as the interacting dumbbell systems.
The time step size is set to be \rev{$\Delta t = 0.01$}, and simulations are 
performed for sufficiently long time. (The stochastic Runge-Kutta scheme
is a weak second order scheme, and this time step is sufficiently small in most cases.)
Simulations are conducted with the house-made code which is available
at the author's web site\cite{Uneyama-website}.
To improve the statistical accuracy, the simulations for the same $\rho$
and $\epsilon$ with different random number seeds are performed and
the averages over different seeds and time are taken.
For the calculation of the relaxation modulus, we employ Likhtman's formula
to further improve the statistical accuracy\cite{Likhtman-chapter}.

%------------------------------------------------------------------------------
\section{RESULTS}
\label{results}

\subsection{Static Properties}
\label{static_properties}

\begin{figure}[tb]
 \centering
 \includegraphics[width=\figurewidth,clip]{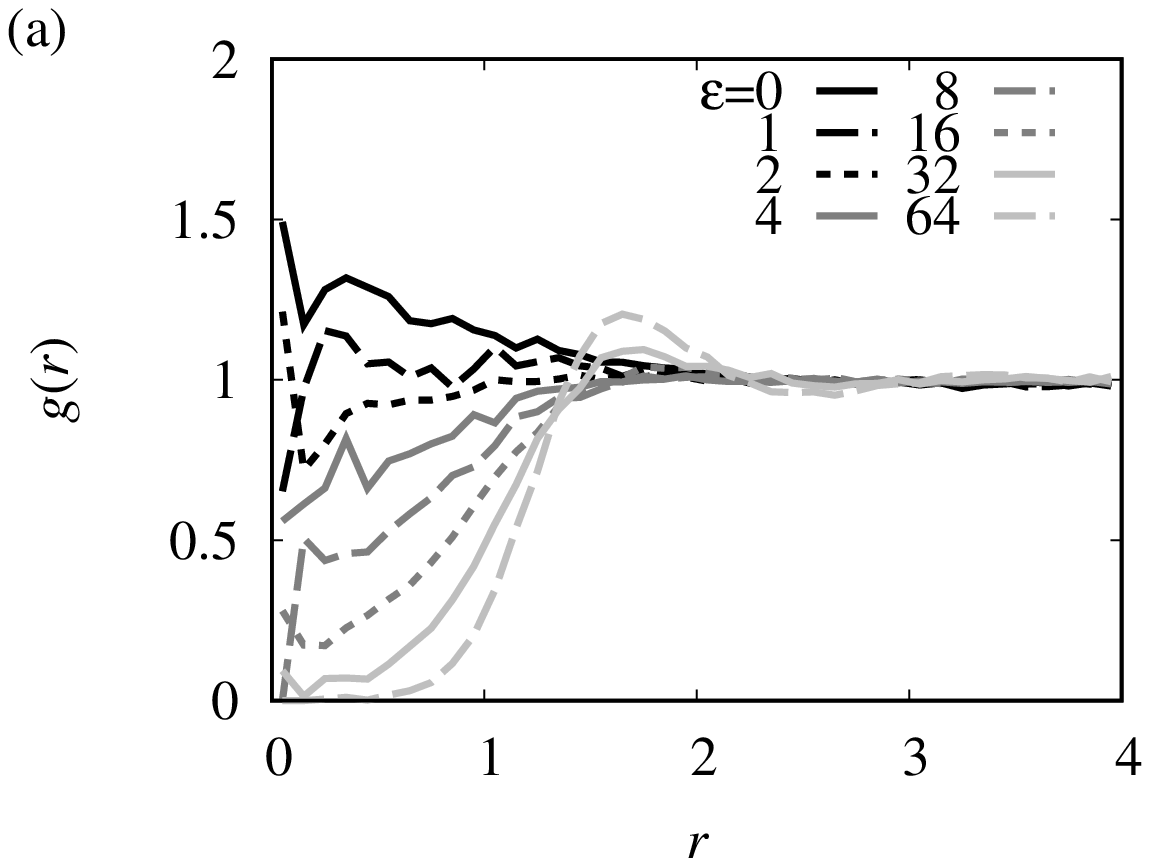}
 \includegraphics[width=\figurewidth,clip]{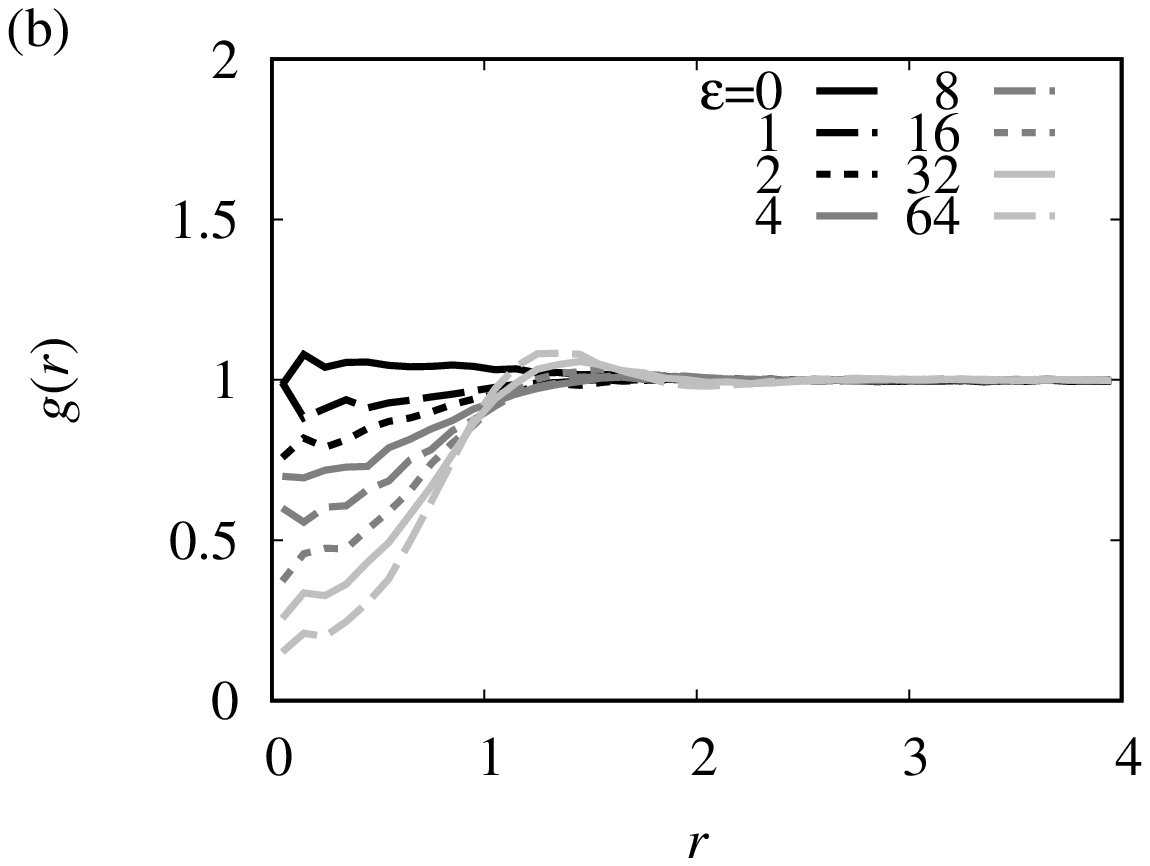}
 \includegraphics[width=\figurewidth,clip]{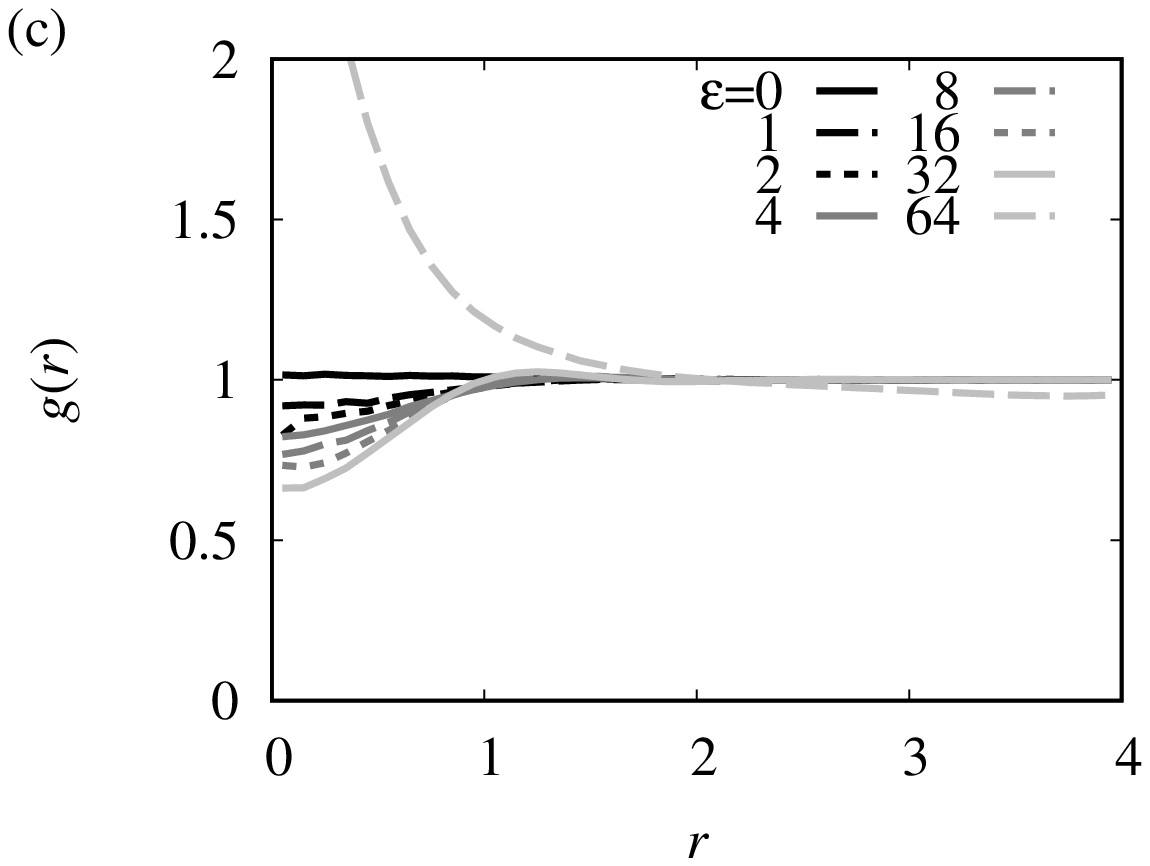}
\caption{\label{rdf_dumbbells}
 The radial distribution functions for the interacting dumbbell systems.
 (a) $\rho = 0.25$, (b) $\rho = 1$, and (c) $\rho = 4$. The data near $r = 0$
 are scattered due to small number of statistical samples.}
\end{figure}

\begin{figure}[tb]
 \centering
 \includegraphics[width=\figurewidth,clip]{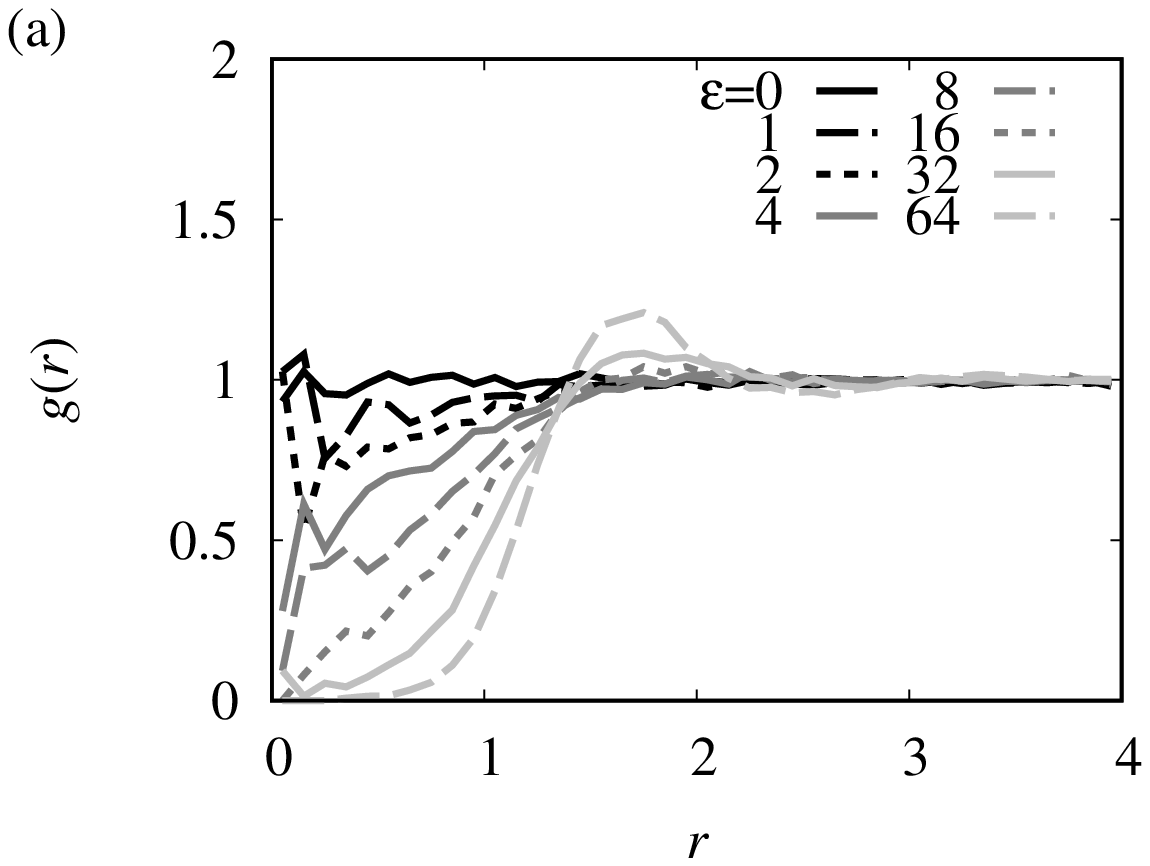}
 \includegraphics[width=\figurewidth,clip]{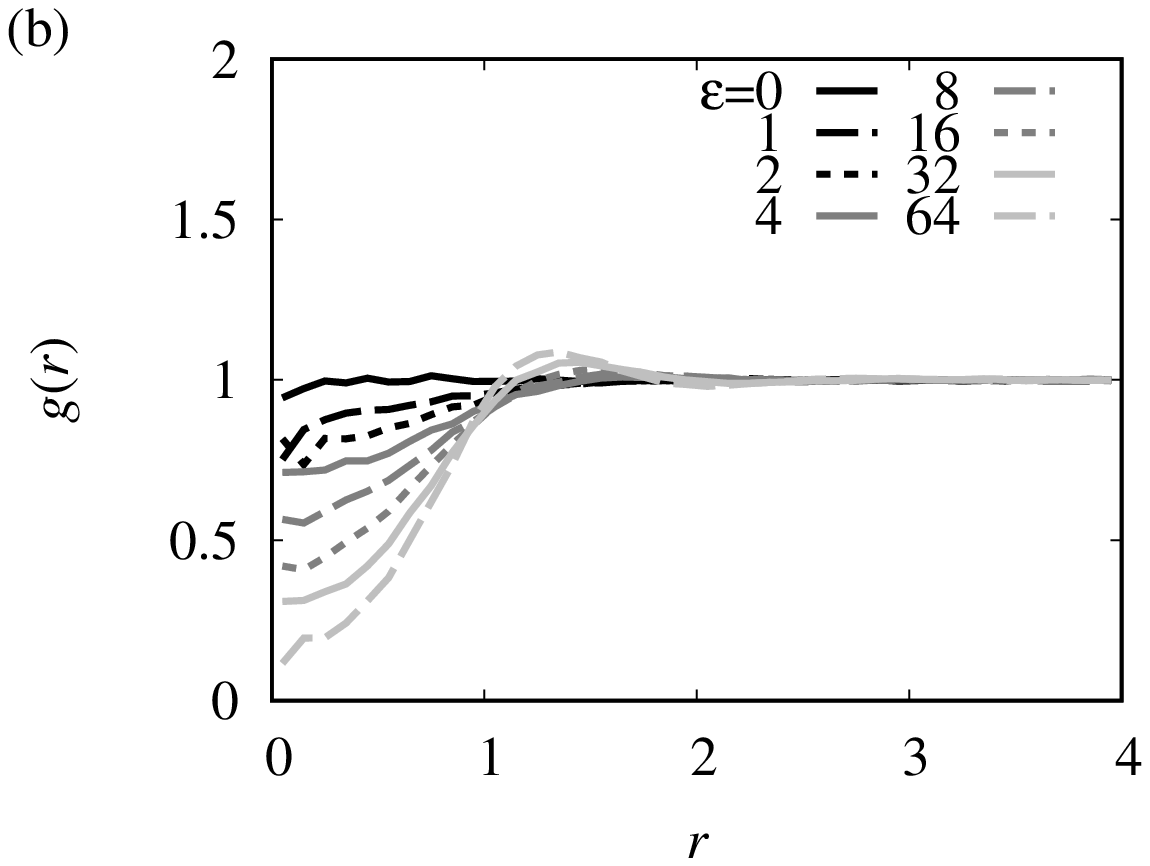}
 \includegraphics[width=\figurewidth,clip]{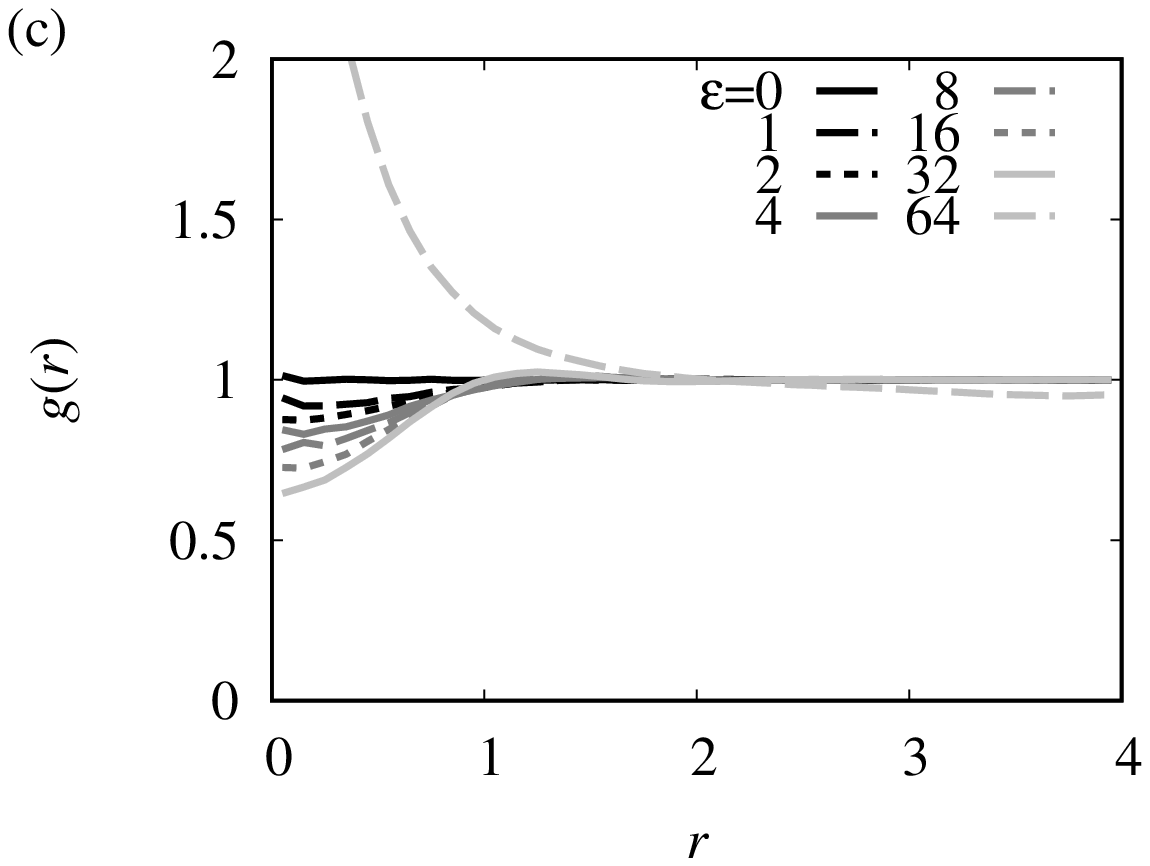}
\caption{\label{rdf_particles}
The radial distribution functions for the soft particle systems.
 (a) $\rho = 0.25$, (b) $\rho = 1$, and (c) $\rho = 4$. Simulation parameters are
 the same as the interacting dumbbell systems shown in Figure~\ref{rdf_dumbbells}.}
\end{figure}

Before we study the linear viscoelasticity, we study some
static properties.
The static structure of the system can be characterized by the radial
distribution function (RDF):
\begin{equation}
 \label{radial_distribution_function}
 g(r) = \frac{1}{4 \pi \rho r^{2}}
  \sum_{(j,k) > (l,m)} \langle \delta(r - |\bm{R}_{j,k} - \bm{R}_{l,m}|) \rangle_{\text{eq}}.
\end{equation}
In a non-interacting ideal dumbbell system, we have $g_{\text{id}}(r) = 1 + (1 / \rho) \exp(-r^{2} / 2)$.
The second term comes from the tethering potential. (In a non-interacting soft particle system, we simply have $g_{\text{id}}(r) = 1$.)
A particle is connected to its partner by the tethering potential,
and should be distributed near $r = 0$.
The contribution of the tethering potential
becomes negligibly small if the density is sufficiently high: $g_{\text{id}}(r) \approx 1$.
The RDF of a system is calculated as a normalized histogram. Notice that
the RDF near $r = 0$ is not accurate, because
the number of samples becomes small for small $r$. (This is not serious
in most of molecular dynamics simulations, because the most of interaction
potentials have hard-core and the RDF becomes zero near $r = 0$. In our
model, however, the soft-core interaction allows particles to overlap
and the RDF can be non-zero even at $r = 0$.)
Figures~\ref{rdf_dumbbells} and \ref{rdf_particles} show the RDFs for interacting dumbbell systems and soft particle systems,
with various densities and interaction parameters.
The RDFs for the interacting dumbbell and soft particle systems with
the same density and interaction parameter are quite similar.
For the interacting dumbbell systems with the low density and low interaction parameter (Figure~\ref{rdf_dumbbells}(a)),
we observe the increase of the RDF near $r = 0$. This is the contribution of the tethering
potential. Except this contribution of the tethering interaction, the
RDF of an interacting dumbbell system are almost the same as that of the corresponding
soft particle system.
Thus we find that the static structure is essentially controlled by
the Gaussian soft-core interaction in these systems.

For the system with the highest density 
and the highest interaction parameter ($\rho = 4$ and $\epsilon = 64$)
among the examined parameter sets,
we observe that the RDF behaves in a qualitatively different way from
other cases. The RDF increases largely as $r$ approaches to zero.
This would be interpreted
as the formation of dimer-like clusters\cite{Miyazaki-Kawasaki-Miyazaki-2016,Miyazaki-Kawasaki-Miyazaki-2019}.
The formation of clusters implies the 
interaction is too strong in this system. We cannot exclude the possiblity
that the simulation model and
numerical scheme employed in this work do not work correctly and have produced
artificial cluster structures. Anyway, we consider the case with
$\rho = 4$ and $\epsilon = 64$ as an exceptional case, and do not directly
compare this case with other cases in what follows.
Except this special case, we observe that the RDF changes systematically
as the density and the interaction parameter changes. The dependence of
the RDF on the interaction parameter becomes weaker as the density
increases. Thus we expect that, at the high density limit, the RDF
will reduce to the ideal form $g(r) \approx g_{\text{id}}(r)$ independent of the interaction
parameter.

\begin{figure}[tb]
 \centering
 \includegraphics[width=\figurewidth,clip]{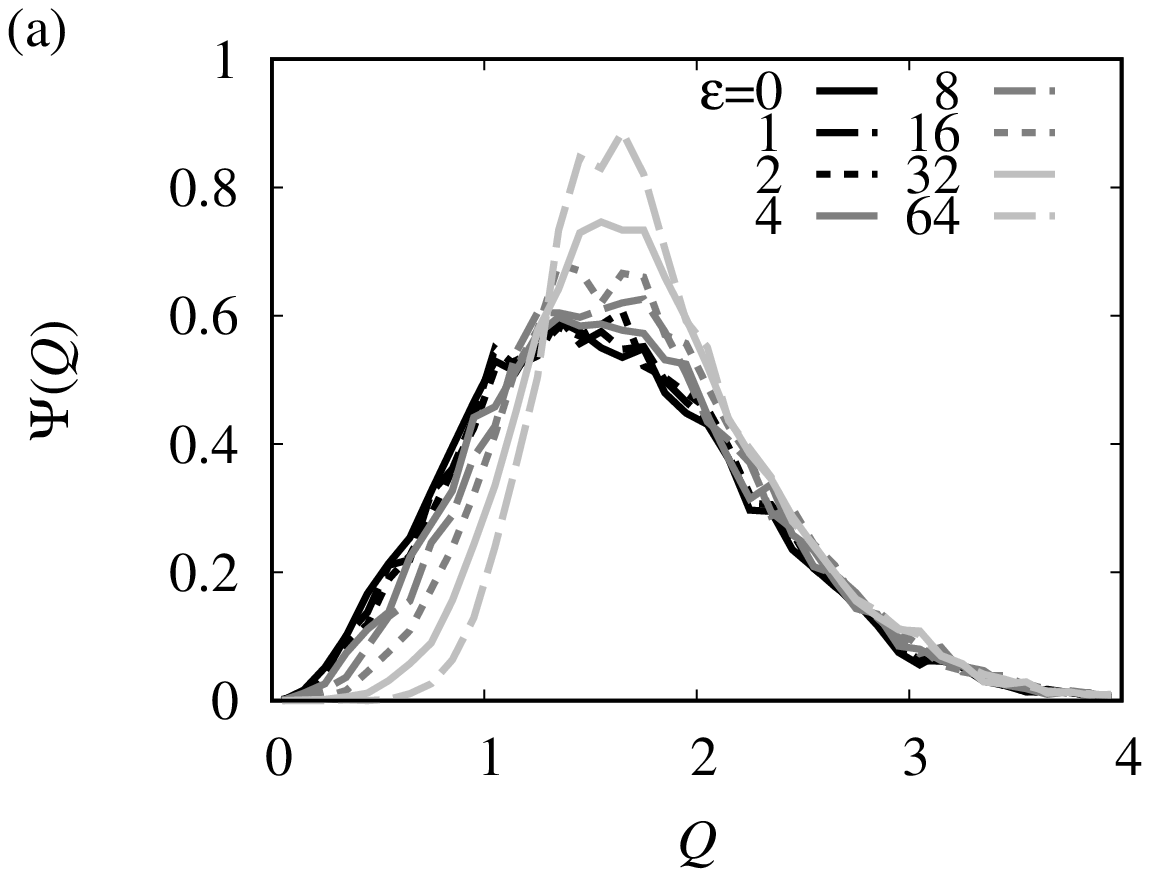}
 \includegraphics[width=\figurewidth,clip]{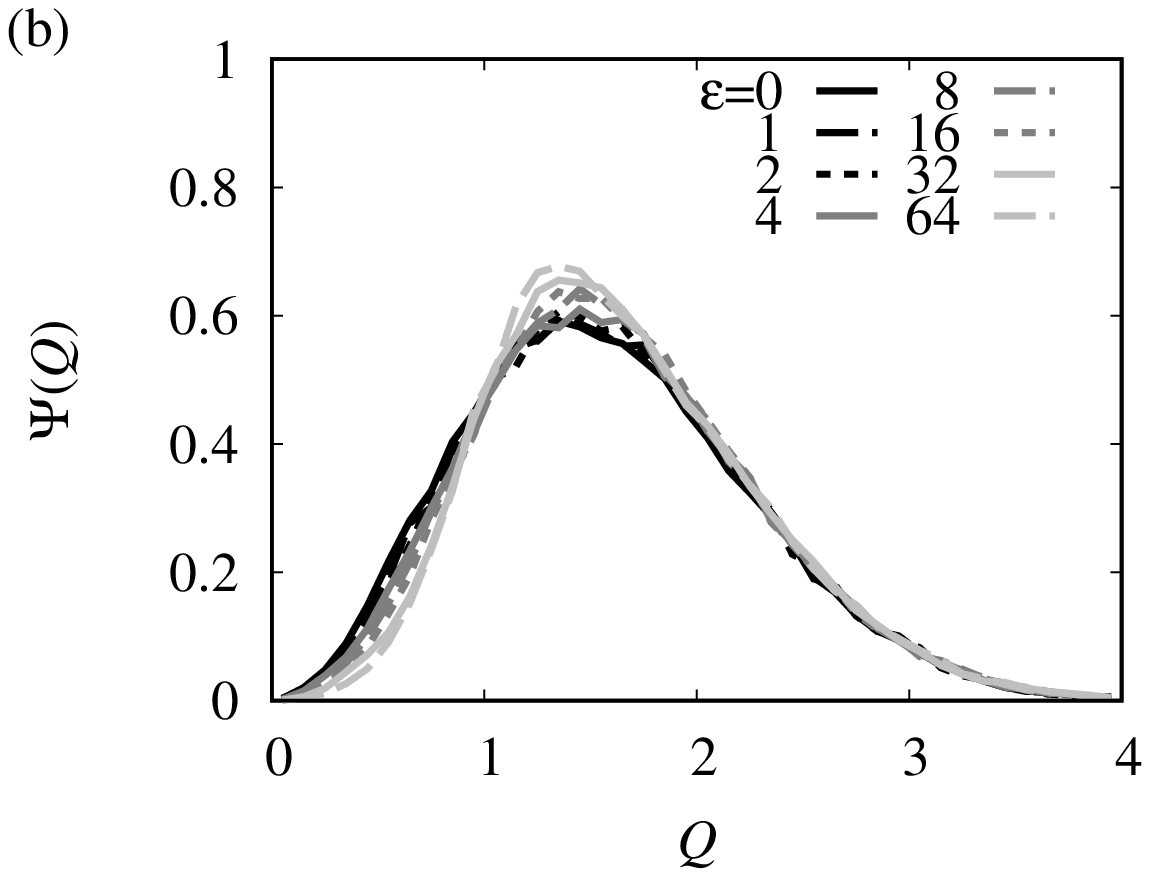}
 \includegraphics[width=\figurewidth,clip]{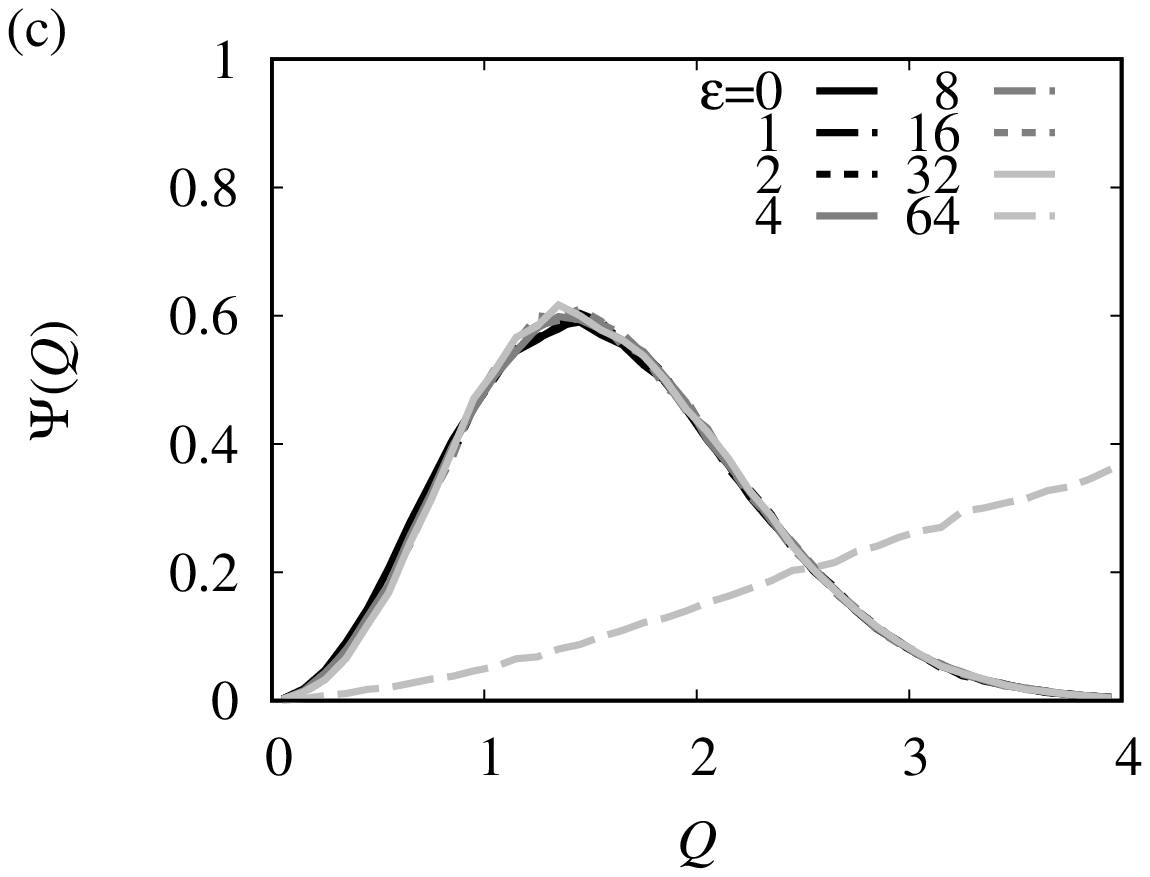}
\caption{\label{bond_dumbbells}
 The bond length distribution functions for the interacting dumbbell systems.
 (a) $\rho = 0.25$, (b) $\rho = 1$, and (c) $\rho = 4$. 
 \rev{Curves for $\epsilon = 0$ (solid black curves) correspond to the distribution function
 for the ideal system $\Psi_{\text{id}}(Q)$.}}
\end{figure}

To study static statistical properties of dumbbell conformations, we
calculate the bond length distribution function. We define the bond length
distribution function as
\begin{equation}
 \label{bond_length_distribution_function}
  \Psi(Q) =
  \frac{1}{N} \sum_{j} \langle \delta(Q - |\bm{R}_{j,1} - \bm{R}_{j,2}|) \rangle_{\text{eq}}.
\end{equation}
In absence of the Gaussian soft-core interaction potential, we simply have
$\Psi_{\text{id}}(Q) = \sqrt{2 / \pi} Q^{2} \exp(-Q^{2} / 2)$. If the equilibrium conformation
is modulated by the Gaussian soft-core interaction, the bond length distribution
should deviate from the equilibrium distribution.
Figure~\ref{bond_dumbbells} shows the bond length distribution functions
for the interacting dumbbell systems.
\rev{(Note that the curves for $\epsilon = 0$ corresponds to the ideal
distribution function $\Psi_{\text{id}}(Q)$.)}
If the density is low, we observe that the bond length distribution
clearly depends on the interaction parameter. 
As the density increases, the dependence of the bond length distribution
on the interaction parameter becomes weak.
The distribution for the exceptional case ($\rho = 4$ and $\epsilon = 64$)
is much different from other distributions as the case of the RDF.
(Judging from this bond length distribution, the systems size $8^{3}$ seems to
be too small in this case. Additional simulations for larger systems should be carefully performed,
if we analyze the data for $\rho = 4$ and $\epsilon = 64$.)
Except this case, we observe that the bond length distribution is
almost independent of $\epsilon$ for the high density systems ($\rho = 4$), and
at the high density limit we will have $\Psi(Q) \approx \Psi_{\text{id}}(Q)$.
These results are consistent with the RDF data.

For low density systems, both the RDF and the bond length distribution function
clearly depends on the interaction
parameter. Such a behavior is widely observed for systems interacting
via such as the Lennard-Jones potential. Our results show that as the density increases, the
distributions become less sensitive to the interaction parameter.
The equilibrium conformational
statistics of a single tagged dumbbell can be well described by that for
the corresponding non-interacting dumbbell.
We may interpret this as the screening effect, because the
RDF will approach to that of the non-interacting ideal system
if the interaction between particles are effectively canceled.
From the RDF and bond length distribution data, therefore
we conclude that the screening effect actually works in our model.

\begin{figure}[tb]
 \centering
 \includegraphics[width=\figurewidth,clip]{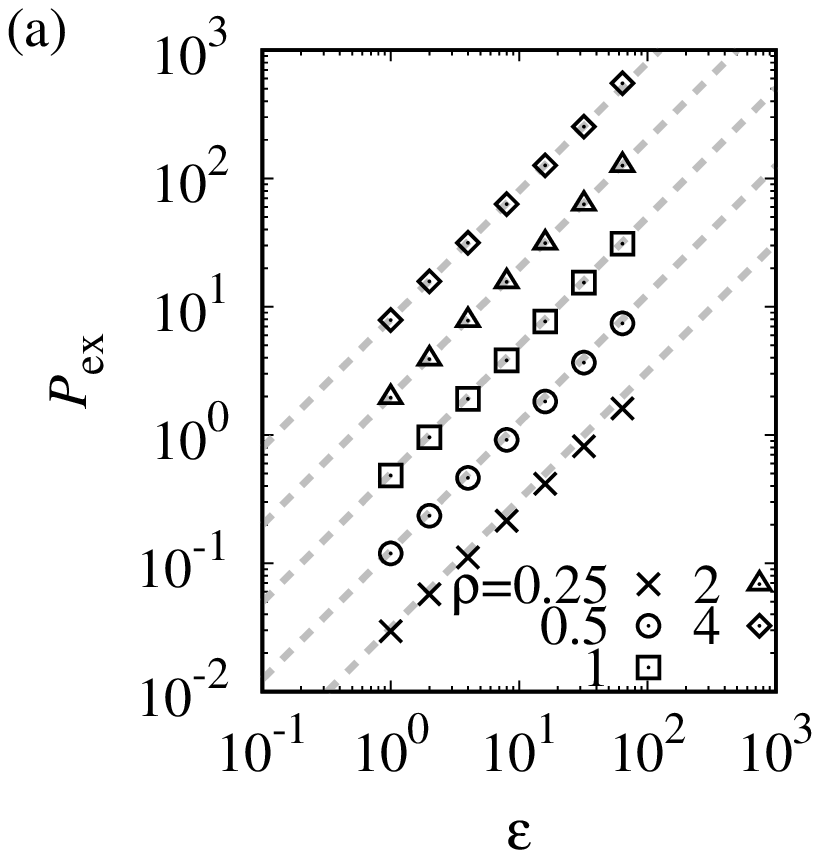}
 \includegraphics[width=\figurewidth,clip]{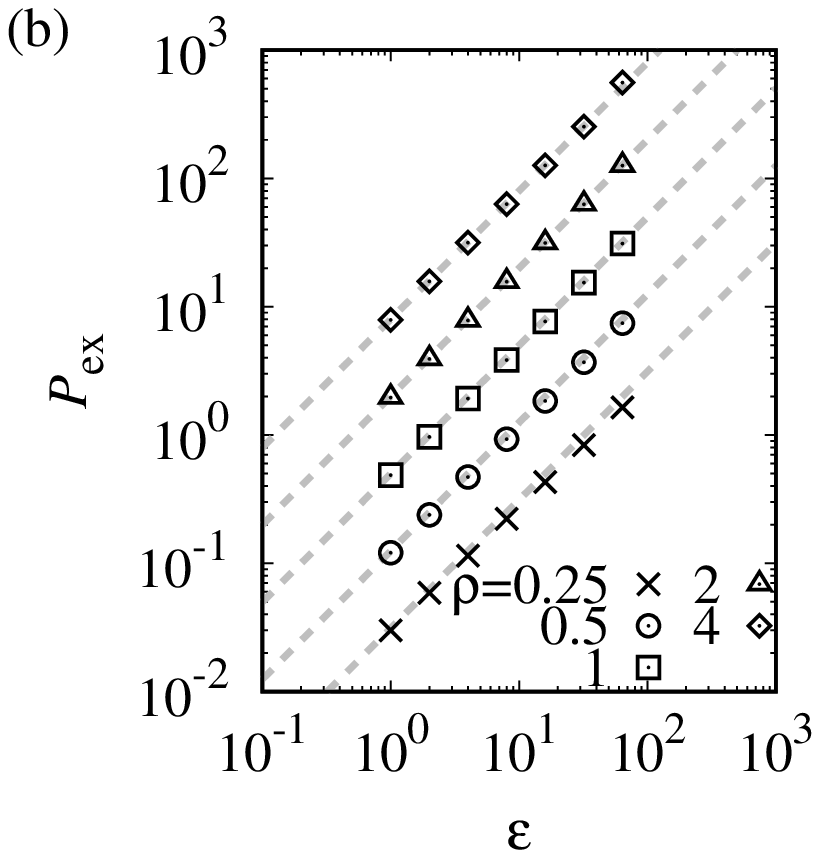}
\caption{\label{pressure_dumbbells_particles}
 The excess pressures calculated by the simulations for (a) the interacting
 dumbbell systems and (b) the soft particle systems, with various densities and
 interaction parameters. Symbols show the simulation data, and the gray dashed
 lines represent the theoretical prediction based on the mean-field approximation (eq~\eqref{excess_pressure_approx}).}
\end{figure}

The Gaussian soft-core interaction between dumbbells would directly affect some
macroscopic thermodynamic quantities. Here we examine the pressure of the system.
\rev{From eq~\eqref{pressure_and_stress}, the pressure of the system can be calculated
from the stress tensor as $P = - (\langle \sigma_{xx} \rangle_{\text{eq}}
+ \langle \sigma_{yy} \rangle_{\text{eq}} + \langle \sigma_{zz} \rangle_{\text{eq}}) / 3$.}
If there is no interaction, the pressure simply obeys the equation of
state for an ideal gas. Thus the pressure is simply given as that of
the ideal gas, $P_{\text{id}} = c \rho $
with $c \rho$ being the number density of molecules in the system.
We have $c = 1 / 2$ and $c = 1$ for the interacting dumbbell
systems and soft particle systems, respectively.
It would be reasonable to subtract this ideal gas pressure $P_{\text{id}}$
from the pressure and define the excess pressure: $P_{\text{ex}} = P - P_{\text{id}}$.
If the target system behaves as an ideal gas, we simply have $P_{\text{ex}} = 0$.
Figure~\ref{pressure_dumbbells_particles} shows the excess pressure of the interacting dumbbell systems and soft
particle systems, \rev{which are calculated from stress tensors}. In Figure~\ref{pressure_dumbbells_particles}, we clearly observe that the excess pressure is non-zero.
As the density $\rho$ and/or the interaction strength $\epsilon$ increases,
the excess pressure increases. 

Louis et al\cite{Louis-Bolhuis-Hansen-2000} and Lang et al\cite{Lang-Likos-Watzlawek-Lowen-2000} analyzed static statistical properties
of the Gaussian soft-core systems. They found that at the high density
limit, the mean-field theory becomes exact. According to their data,
the mean-field theory seems to be accurate even if the density is not
very high. Lang et al derived a simple approximate expression for the excess pressure\cite{Lang-Likos-Watzlawek-Lowen-2000}:
\begin{equation}
 \label{excess_pressure_approx}
 P_{\text{ex}} \approx \frac{1}{2} \epsilon \rho^{2} .
\end{equation}
(Notice that the definition of the interaction parameter $\epsilon$ in Ref.~\cite{Lang-Likos-Watzlawek-Lowen-2000} is different
from that in this work.)
The prediction by eq~\eqref{excess_pressure_approx} is shown in Figure~\ref{pressure_dumbbells_particles}
as gray dashed lines. The agreement between the simulation data and
the theoretical prediction is very good even at the low density systems ($\rho = 0.25$).
Thus we find that the mean-field approximation works very well
for the interacting dumbbell systems and soft particle systems examined in this work.

The static simulation data would be summarized as follows.
From the structural data, we found that the structure of the interacting dumbbells
approach to that of the corresponding non-interacting system. However,
the pressure data show that the dumbbells are strongly interacting while
the mean-field approximation works very well. In some aspects, the Gaussian
soft-core interaction seems to be canceled and can be ignored,
but in some other aspects, it cannot be simply ignored.

\subsection{Linear Viscoelasticity}
\label{linear_viscoelasticity}

\begin{figure}[tb]
 \centering
 \includegraphics[width=\figurewidth,clip]{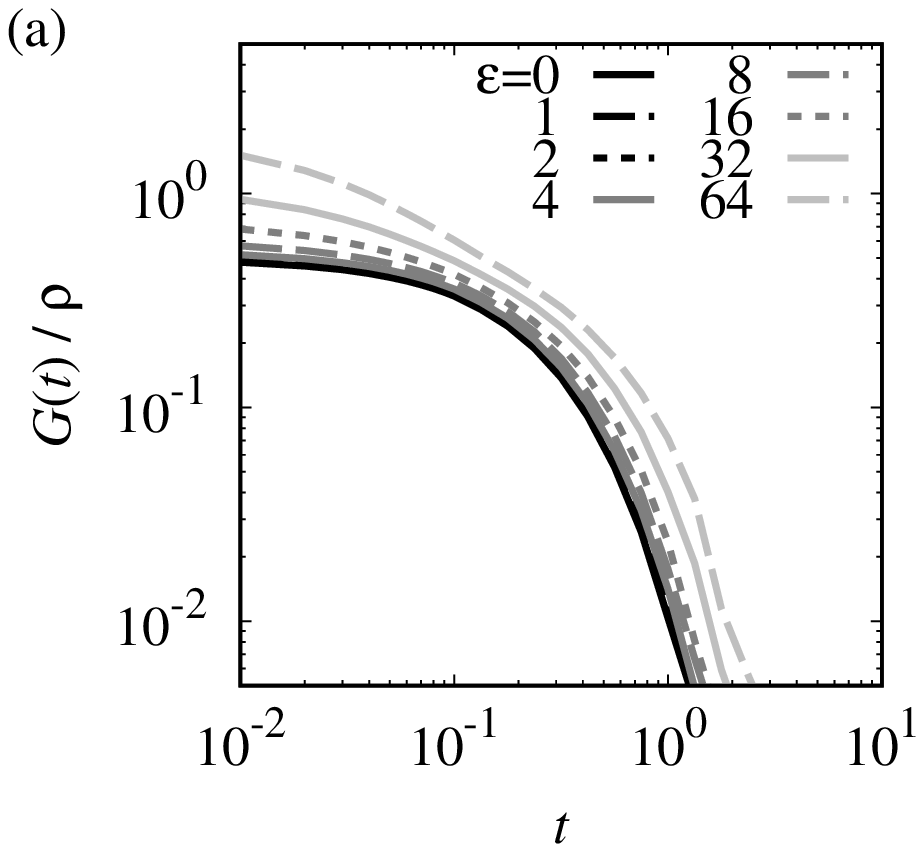}
 \includegraphics[width=\figurewidth,clip]{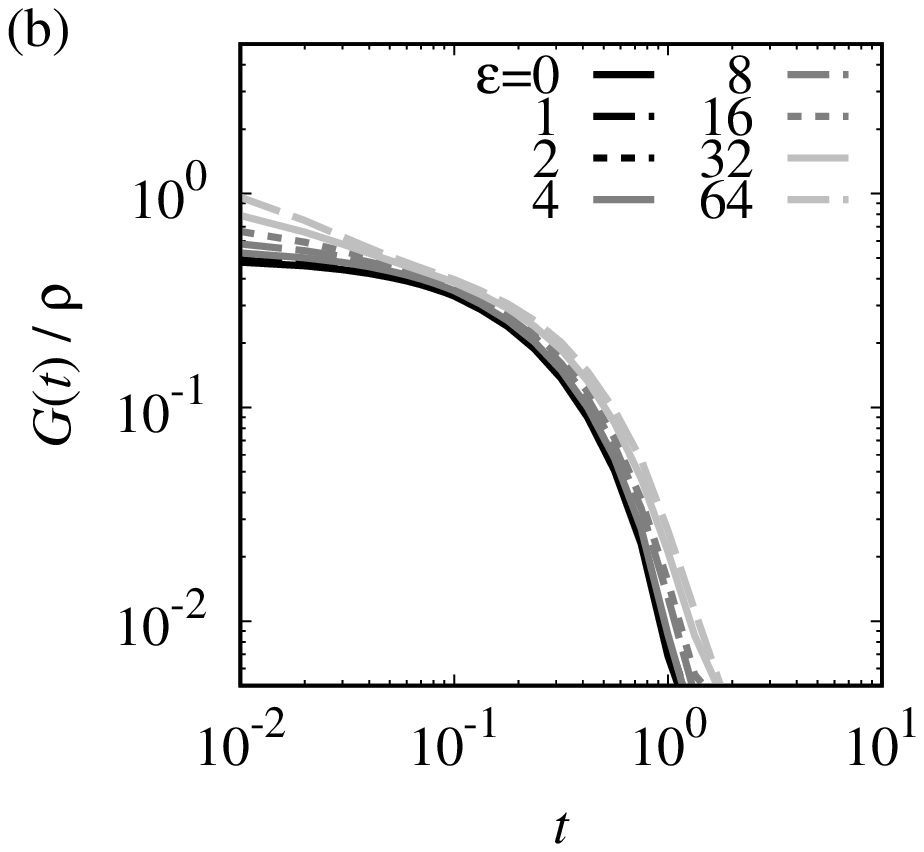}
 \includegraphics[width=\figurewidth,clip]{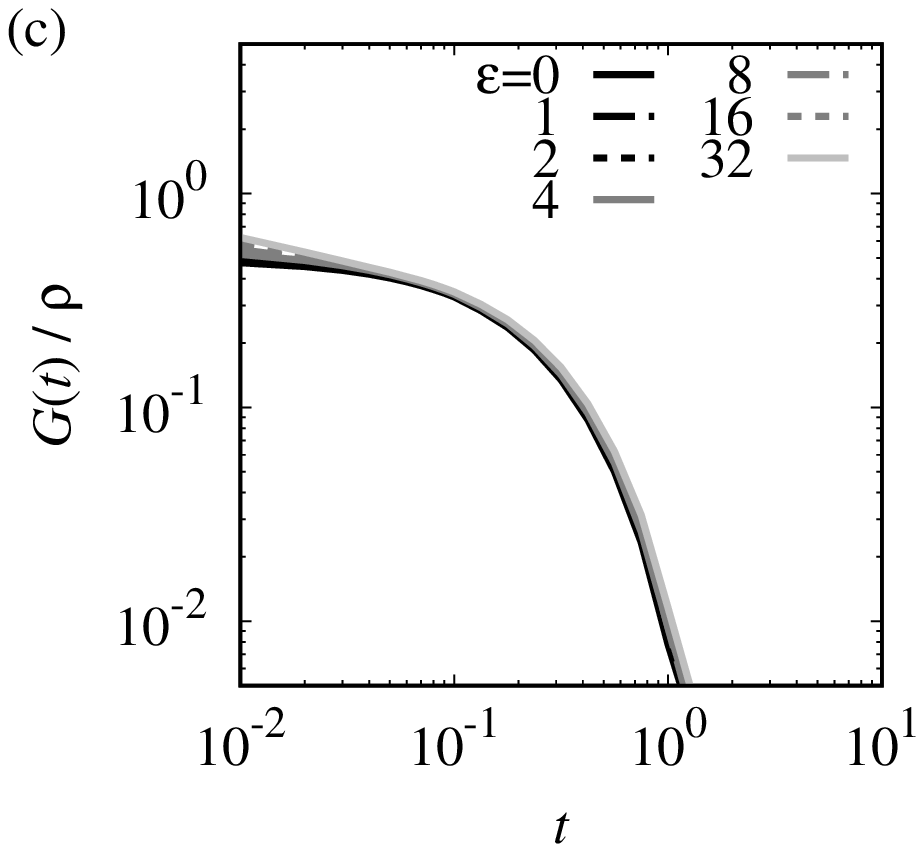}
\caption{\label{relaxation_modulus_dumbbells}
 The shear relaxation moduli for the interacting dumbbell systems.
 (a) $\rho = 0.25$, (b) $\rho = 1$, and (c) $\rho = 4$. The relaxation modulus
 for the system with $\rho = 4$ and $\epsilon = 64$ is much larger than other cases and not shown here.
 \rev{Curves for $\epsilon = 0$ (solid black curves) correspond to the relaxation modulus
 for the ideal system $G_{\text{id}}(t) / \rho$.}}
\end{figure}

From the static data, we cannot judge whether the dynamics of
interacting dumbbells can be reasonably approximated by that of
non-interacting dumbbells. Judging from the fact that the pressure is not screened,
the linear viscoelasticitic properties may also reflect the strong interaction between particles.
In the case of ideal non-interacting dumbbells ($\epsilon = 0$), 
from eqs~\eqref{stress_tensor} and \eqref{green_kubo_relation},
we simply have
\begin{equation}
 \label{relaxation_modulus_ideal}
 G(t) = G_{\text{id}}(t) = \frac{\rho}{2} e^{-4 t}.
\end{equation}
\rev{Judging from the static structure data, we expect that the ideal
relaxation modulus may be reproduced if the interaction is screened.
Thus, by comparing the relaxation moduli of interacting dumbbells with
the ideal relaxation modulus, we will be able to judge the screening from
the view point of dynamics. (If the interaction in an interacting dumbbell system
is actually screened, we expect that $G(t)$ of that system simply reduces to the ideal one, as eq~\eqref{relaxation_modulus_ideal}.)}
It would be convenient to use the reduced relaxation modulus
$G(t) / \rho$.
We show the reduced relaxation moduli of the interacting dumbbell systems with
various densities and interaction parameters in Figure~\ref{relaxation_modulus_dumbbells}.
\rev{(Note that the curves for $\epsilon = 0$ correspond to the relaxation modulus of the ideal system, $G_{\text{id}}(t) / \rho = e^{-4 t} / 2$.)}
In the exceptional case $\rho = 4$ and $\epsilon = 64$ (data not shown),
the relaxation modulus seems to have very long relaxation time. This might be
due to the glassy dynamics\cite{Miyazaki-Kawasaki-Miyazaki-2016,Miyazaki-Kawasaki-Miyazaki-2019},
although we do not go into the detail in this work.
If the density is relatively high, we observe that the relaxation moduli depend on the
interaction parameter weekly and eq~\eqref{relaxation_modulus_ideal} works as
a reasonable approximation. If the density is not high, however, we observe that
the reduced shear relation modulus clearly depends on the interaction parameter.
For example, in Figure~\ref{relaxation_modulus_dumbbells}(a), the relaxation modulus
for the case of $\epsilon = 64$ is much different from eq~\eqref{relaxation_modulus_ideal}.
The longest relaxation time of the system with $\rho = 0.25$ and $\epsilon = 64$ is slightly
longer than the ideal one. Also, the relaxation modulus has broader relaxation
time distribution. The relaxation
behavior at the short time region ($10^{-2} \lesssim t \lesssim 10^{-1}$) looks rather
close to the power-law type relaxation than the single exponential relaxation.

\begin{figure}[tb]
 \centering
 \includegraphics[width=\figurewidth,clip]{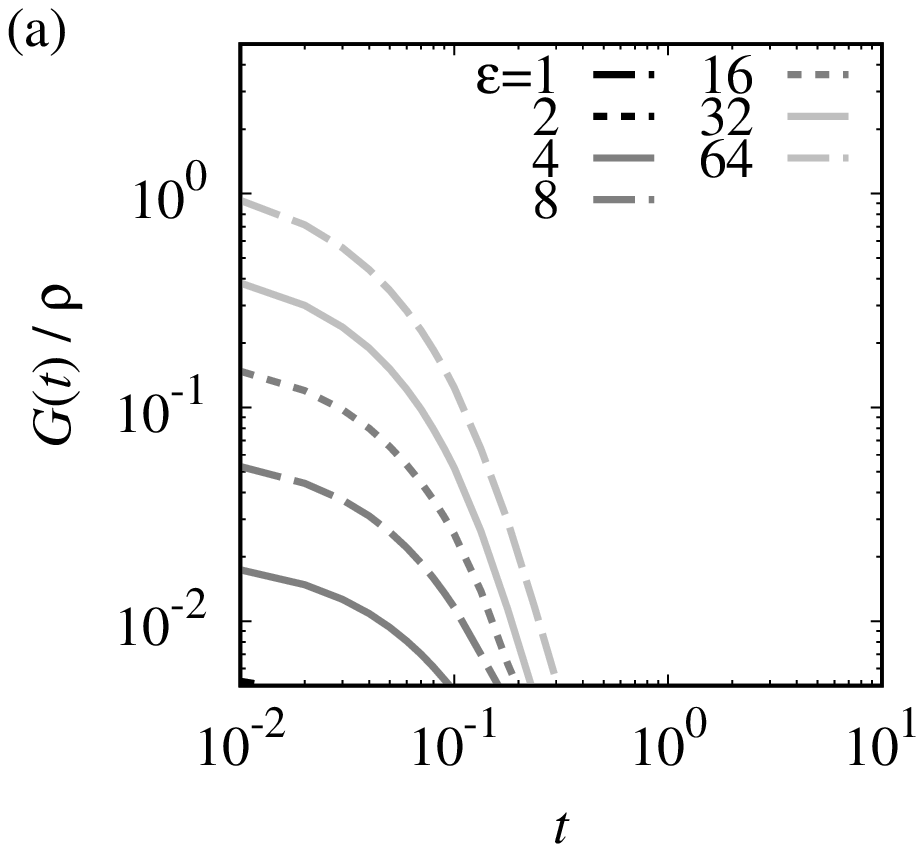}
 \includegraphics[width=\figurewidth,clip]{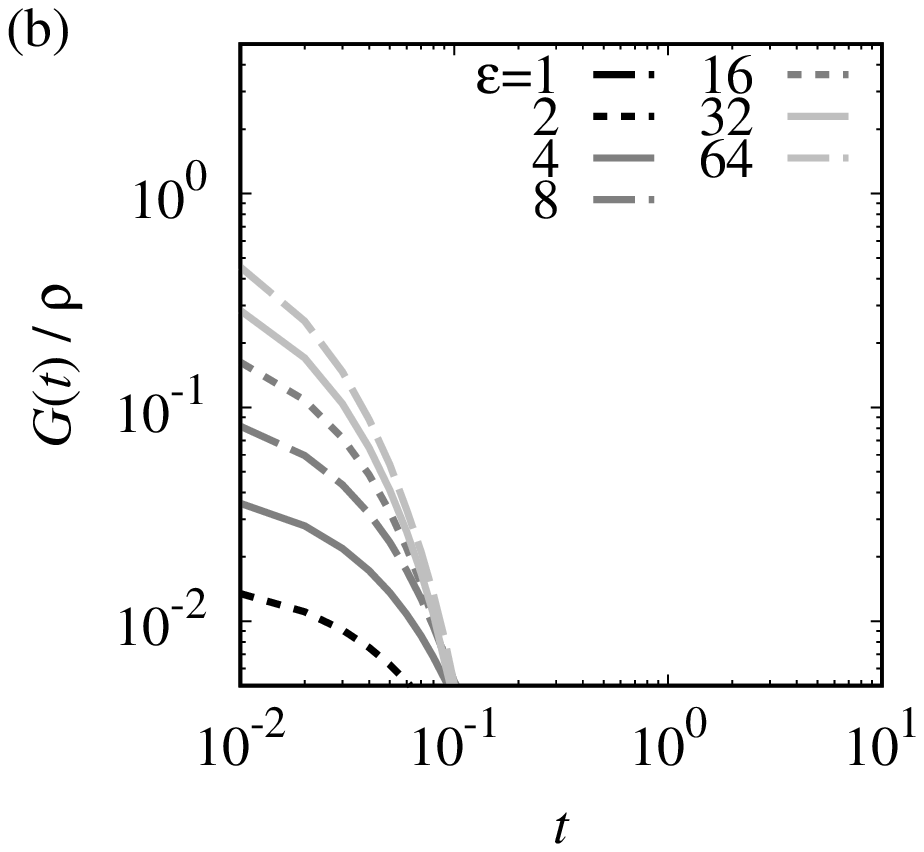}
 \includegraphics[width=\figurewidth,clip]{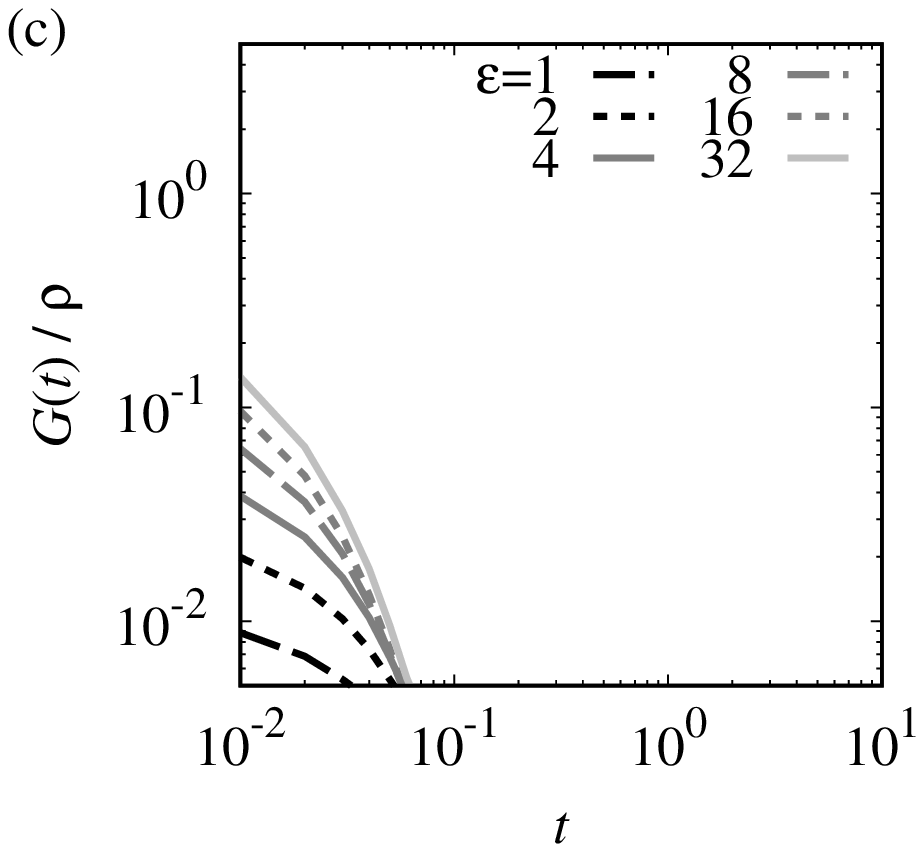}
\caption{\label{relaxation_modulus_particles}
 The shear relaxation moduli for the soft particle systems.
 (a) $\rho = 0.25$, (b) $\rho = 1$, and (c) $\rho = 4$. The data for $\epsilon = 0$ are not shown,
 since they are trivially zero. Also, the relaxation modulus
 for the system with $\rho = 4$ and $\epsilon = 64$ is not shown. The relaxation moduli for relatively small $\epsilon$
 may not be observed in the presented range.
 Simulation parameters are
 the same as the interacting dumbbell systems shown in Figure~\ref{relaxation_modulus_dumbbells}.}
\end{figure}

We should recall that the relaxation moduli shown in Figure~\ref{relaxation_modulus_dumbbells}
contain contributions of the tethering potential and the Gaussian soft-core potential.
Therefore, the relaxation behavior at the short-time region may be attributed to
the Gaussian soft-core interaction. Figure~\ref{relaxation_modulus_particles}
shows the shear relaxation moduli for the soft particle systems. The densities
and interaction parameters are the same as those in Figure~\ref{relaxation_modulus_dumbbells}.
The soft particle systems relax around $t \approx 10^{-1}$.
If the density is high, the reduced relaxation moduli of soft particle systems are
small. Thus the contribution of the Gaussian soft-core interaction will be
negligibly small in the high-density interacting dumbbell systems.
If the density is low, the reduced relaxation moduli of soft particle systems
are rather large.
Thus we expect that the relaxation modes due to Gaussian soft-core interaction
can be observed at the short-time region in the interacting dumbbell model, if
the density is low. The short-time relaxation modes can
be simply attributed to the Gaussian soft-core interaction.

\begin{figure}[tb]
 \centering
 \includegraphics[width=\figurewidth,clip]{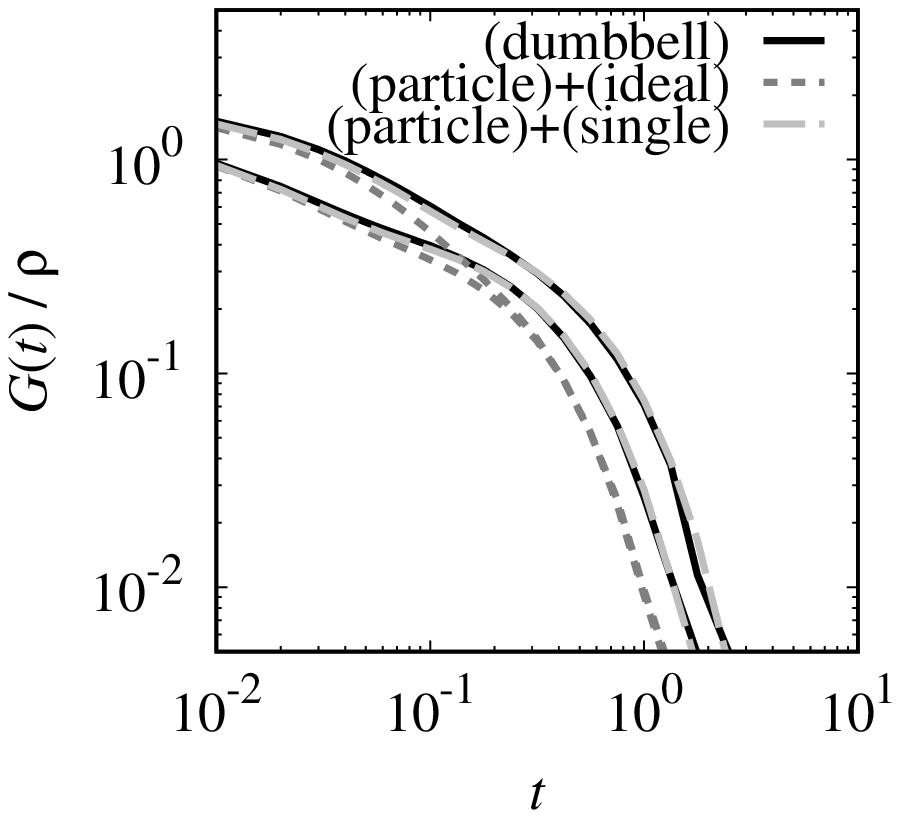}
\caption{\label{relaxation_modulus_reconstruction}
 The shear relaxation moduli for the soft dumbbell systems
 and the reconstructed shear relaxation moduli. The density and interaction parameter
 are set as $\rho = 0.25$ and $1$, and $\epsilon = 64$, respectively.
 The solid black curves represent the simulation data (taken from Figure~\ref{relaxation_modulus_dumbbells}).
 The dark gray dotted curves and light gray dashed curves represent
 the reconstruction results based on the relaxation moduli for the soft particle systems.
 The dark gray dotted curves and 
 light gray dashed curves utilize the ideal relaxation modulus (eq~\eqref{relaxation_modulus_ideal})
 and the single-mode Maxwell model with empirical parameters (eq\eqref{relaxation_modulus_single}), respectively.}
\end{figure}

Even if the short-time relaxation modes are attributed to the Gaussian soft-core
interaction, the relaxation modes of the tethering potential
in the interacting dumbbell systems are
generally different from the ideal form (eq~\eqref{relaxation_modulus_ideal}).
If the bond relaxation mode (the relaxation mode of the tethering potential) is not affected by the
Gaussian soft-core interaction, we expect that the relaxation modulus of
an interacting dumbbell system can be reconstructed as
$G_{\text{dumbbells}}(t) = G_{\text{particles}}(t) + G_{\text{id}}(t)$ with
$G_{\text{particles}}(t)$ being the relaxation modulus of the corresponding
soft particle system. This reconstruction does not work if the relaxation mode
of the tethering interaction is affected by the Gaussian soft-core interaction.
If we assume that the effect of the Gaussian soft-core interaction to this
relaxation mode can be expressed as the modulation of the relaxation intensity
and the relaxation time, we may employ the following empirical form for the relaxation modulus:
\begin{align}
 \label{relaxation_modulus_reconstruction_with_single}
 G_{\text{dumbbells}}(t) & = G_{\text{particles}}(t) + G_{\text{single}}(t), \\
 \label{relaxation_modulus_single}
 G_{\text{single}}(t) & = \frac{\rho}{2} \psi_{\text{eff}} e^{-t / \tau_{\text{eff}}}.
\end{align}
Here, $G_{\text{single}}(t)$ is the single-mode Maxwell relaxation, and
$\psi_{\text{eff}}$ and $\tau_{\text{eff}}$ represent the effective interaction
strength and the effective relaxation time, respectively.
(In the ideal case where the Gaussian soft-core interaction is absence, we have 
$\psi_{\text{eff}} = 1$
and $\tau_{\text{eff}} = 1 / 4$ from $G_{\text{single}}(t) = G_{\text{id}}(t)$.)
These parameters should
be determined so that the reconstructed relaxation modulus by eq~\eqref{relaxation_modulus_reconstruction_with_single}
agrees with the simulation data for the interacting dumbbell systems.
We show the relaxation moduli by these reconstruction
methods together with the simulation data for some interacting dumbbell systems
in Figure~\ref{relaxation_modulus_reconstruction}.
The effective relaxation strengths and relaxation times are set as $(\psi_{\text{eff}},\tau_{\text{eff}}) = (1.1,0.50)$ for
$(\rho, \epsilon) = (0.25, 64)$, and $(\psi_{\text{eff}},\tau_{\text{eff}}) = (1.0,0.35)$ for $(\rho,\epsilon) = (1,64)$.
We observe that the reconstruction with the ideal relaxation modulus does not work
while the reconstruction by eq~\eqref{relaxation_modulus_reconstruction_with_single} works almost perfectly.
From this result, we conclude that the short-time relaxation modes are caused
by the Gaussian soft-core interaction, and the long-time bond relaxation mode
can be well described by a single-mode Maxwell model.
The effective relaxation time becomes slightly longer than that for the ideal system.

%------------------------------------------------------------------------------

\section{DISCUSSIONS}
\label{discussions}

Although the interaction between dumbbells seem to have rather
weak contribution to the shear relaxation modulus from the simulation data
(Figures~\ref{relaxation_modulus_dumbbells} and \ref{relaxation_modulus_reconstruction}),
we should recall that the dumbbells are strongly interacting each other. This is
clear from the pressure data (Figure~\ref{pressure_dumbbells_particles}).
The simulation data for the excess pressure
simply mean that an interacting dumbbell system (or a soft particle system)
can never be treated as an ideal gas. This is not surprising, because we
know that polymer chains behave as ideal Gaussian chains in a polymer melt,
and at the same time, a polymer melt is almost incompressible. 
From eq~\eqref{excess_pressure_approx}, the compressibility of the system
is calculated as
\begin{equation}
 \label{compressibility}
  \kappa = \frac{1}{\rho (\partial P / \partial \rho)}
  \approx  \frac{1}{c \rho + \epsilon \rho^{2}}.
\end{equation}
As the density and/or the interaction parameter increases, the compressibility
decreases. In most of polymer melts
the interaction is sufficiently strong and the compressibility is sufficiently
low. Therefore, we consider that the interacting dumbbell model will be
utilized as a simple model for polymer melts, when the density and/or the
interaction parameter is high. 

As we observed, the shear relaxation moduli
of the interacting dumbbells at high density are close to those of
non-interacting dumbbells (Figure~\ref{relaxation_modulus_dumbbells}(c)).
A single dumbbell in the system behaves in almost the same way as an
non-interacting ideal dumbbell. The situation will not be changed even if
we replace interacting dumbbells by interacting Rouse chains.
A single Rouse chain in such a system will behave as a non-interacting ideal chain,
if the density is sufficiently high.
This empirically supports the naive picture of the
screening and the application of the Rouse model to unentangled polymer melts.

However, the consideration above implies that the validity of the screening
and the recovery of the ideal Rouse dynamics depend on the scale.
Here we consider
the coarse-graining of a simple Rouse type chain to estimate the scale-dependence
of the screening effect. We model a single polymer
chain by connecting $N_{0}$ segments by harmonic springs. We assume that segments
are already coarse-grained and the interaction between segments can be
approximately expressed by a Gaussian soft-core potential.
We express the segment size as $b_{0}$
and the segment density as $\rho_{0}$. We define a coarse-grained
segment which consists of $\alpha$ segments. Then the polymer chain is
reinterpreted as a chain in which $N_{1}$ segments are
connected by harmonic springs. The coarse-grained number of segments and segment size
are $N_{1} = \alpha^{-1} N_{0}$ and $b_{1} = \alpha^{1/2} b_{0}$.
The segment density is changed as $\rho_{1} = \alpha^{-1} \rho_{0}$.
From the viewpoint of the interacting dumbbell and soft particle models,
we should use the dimensionless segment density. For the models before and
after the coarse-graining, we have $\rho_{0} b_{0}^{3}$ and $\rho_{1} b_{1}^{3} = \alpha^{1/2} (\rho_{0} b_{0}^{3})$.
Therefore, by the coarse-graining, the dimensionless segment density increases
by the factor $\alpha^{1/2}$.
The interaction parameter is also changed by this coarse-graining.
We express the interaction parameter before and after the coarse-graining
as $\epsilon_{0}$ and $\epsilon_{1}$. Under the mean-field approximation,
the interaction energy per unit volume should be invariant under the
coarse-graining: $\epsilon_{0} \rho_{0}^{2} / 2 = \epsilon_{1} \rho_{1}^{2} / 2$.
Thus we have $\epsilon_{1} = \alpha^{2} \epsilon_{0}$.
We should use the dimensionless unit to discuss the effective interaction strength.
The dimensionless interaction parameters are roughly estimated to be
$\epsilon_{0} / k_{B} T b_{0}^{3}$ and
$\epsilon_{1} / k_{B} T b_{1}^{3}$, and then we have
$\epsilon_{1} / k_{B} T b_{1}^{3} = \alpha^{1/2} (\epsilon_{0} / k_{B} T b_{0}^{3})$.
This means that the dimensionless interaction parameter increases by the factor $\alpha^{1/2}$ by the coarse-graining.

From the estimates shown above, both the dimensionless segment density
and the dimensionless interaction parameter change as the coarse-graining
is performed.
If the coarse-graining is performed further, the dimensionless segment density
increases further, and the mean-filed description for the Gaussian soft-core
interaction becomes more accurate. Although the interaction parameter also increases
with the coarse-graining, the relaxation modulus is not sensitive to the interaction parameter
(as far as the dimer-like clusters are not formed).
As a result, at a high coarse-graining
level, the screening effect is safely justified. The lower
order Rouse modes corresponds to such highly coarse-grained dynamics, and
such Rouse modes would behave in almost the
same way as in ideal non-interacting systems.
This result would also justify some coarse-grained multi-chain models which are based on the
non-interacting ideal chain statistics\cite{Uneyama-Masubuchi-2012,Uneyama-2019}.

On the other hand, at relatively small scales where the coarse-graining level is low,
the screening is fully not justified. Thus the higher order Rouse modes may
deviate from the ideal behavior. We should recall that we observed
the relaxation modes by the Gaussian soft-core interaction at the short-time
scale region. The Rouse relaxation modes will be mixed-up with the
Gaussian soft-core interaction modes. For the case of the dumbbell model,
we observed that the relaxation modulus apparently looks power-law like if
the interaction is not screened. If we naively consider the situation is
similar for the higher order Rouse modes, we will observe the sum of
power-law like relaxations at the short-time region. The sum of
Rouse relaxation modes itself is already a power-law like relaxation (with the 
power-law exponent $1/2$). Then, even if the Gaussian soft-core interaction
modulates a Rouse relaxation mode from a single exponential form to a power-law form,
the resulting relaxation modulus may just look like a power-law relaxation.
Therefore we expect that we may still observe the Rouse-like 
power-law relaxation at the short time region, even if the screening does not work and the contribution
of the Gaussian soft-core interaction remains.
Only from the Rouse-like functional form of the relaxation modulus,
we may not be able to judge whether the interaction between the segments is really
screened or not. We may need to carefully analyze several dynamical
quantities to judge the screening effect experimentally. Conversely,
when we study the linear viscoelastic properties, the screening effect
may be naively assumed (because the results will not be sensitive to the
screening effect).

\rev{Another example where the screening may not be fully justified is a polymer solution
(at the theta point).
In the polymer solution, the effective density of polymer chains can be rather low.
Then the screening effect may be insufficient and the bond length distribution and 
the relaxation modulus may deviate from the ideal ones. Then, even for lower order
relaxation modes, we may observe the deviation from the ideal behavior.
However, we should recall that there is the hydrodynamic interaction in a polymer solution.
The hydrodynamic interaction may not be screened, when the interaction potential is
not fully screened. Then the relaxation modulus will be affected by both
the hydrodynamic interaction and the interaction potential. The 
interaction strongly affects the relaxation modulus (as predicted by the
Zimm model), and thus is primarily important.
The effect of the interaction potential may be secondary and may not be
clearly observed even if it exists.}

Before we end this section, we should recall that we have examined the screening effect only in
equilibrium. This means that we cannot discuss the nonlinear rheological behavior
based on our simulation results. Hess considered an interacting dumbbell system
and proposed a constitutive equation model\cite{Hess-1984}. Hess constructed
an effective dynamics model for a single dumbbell under the mean-filed approximation.
The effects of the interaction between dumbbells is expressed as the 
effective tethering interaction potential with the
effective interaction matrix. This effective interaction matrix reflects
the average conformational change, and thus we observe nonlinear rheological
behavior such as the shear thinning behavior.
From the view point of nonequilibrium statistical
mechanics, it would be reasonable to introduce the mean-field
effective mobility\cite{Uneyama-Horio-Watanabe-2011} instead of the effective tethering potential.
In this picture, the mobility (or the friction coefficient) is modeled as
an anistropic tensorial quantity which reflects the average conformational change.
In both cases, 
we observe that the rheological behavior is largely changed under fast shear flows.
This is consistent with the experimental
fact that unentangled polymer melts exhibit shear thinning behavior
which cannot be described by the Rouse model.
We will need to perform additional simulations
under nonequilibrium conditions, to study how the screening effect works
in nonequilibrium states and how we can model the nonequilibrium dynamics. It is beyond the scope of the current work and
will be an interesting future work.

%------------------------------------------------------------------------------
\section{CONCLUSIONS}

We have performed simulations for interacting soft dumbbell systems and
soft particle systems, to study the screening effect. From the static
structural data (the RDFs and the bond length distributions), we found
that the screening reasonably works if the density is high. This does not
mean that an interacting dumbbell system becomes approximately equivalent
to the corresponding non-interacting ideal dumbbell system.
The pressure increases as the density increases, and thus an interacting
dumbbell system becomes less compressible as the density increases.

The relaxation moduli of the interacting dumbbell exhibit similar screening
effect. For high density systems, the relaxation moduli are very close to the
relaxation moduli of the non-interacting ideal dumbbell systems. This result
means that the screening effect works both for static structures and
dynamics of interacting dumbbells. Our simulation results would partly
support the usage of the naive screening picture in various coarse-grained
dynamics model for polymer melts. If the density is low, we observe
the relaxation modulus deviates from the ideal one. However, the
short-time relaxation modes observed in low density systems can be attributed
to the relaxation modes of the Gaussian soft-core interaction. The bond
relaxation mode can be expressed as a single Maxwell relaxation mode, although
the relaxation intensity and relaxation time are slightly modulated from ideal ones.
If we assume that the situation is similar in the case of the interacting Rouse chains,
we expect that the relaxation modulus of the interacting Rouse model will be
indistinguishably similar to the ideal Rouse model without interaction between segments.

%------------------------------------------------------------------------------
\section*{ACKNOWLEDGMENT}

This work was supported by JST, PRESTO Grant Number JPMJPR1992, Japan,
and Grant-in-Aid (KAKENHI) for Transformative Research Areas B JP20H05736.

%------------------------------------------------------------------------------
% reference
%\bibliographystyle{srj}
%\bibliography{interacting_dumbbells}

%------------------------------------------------------------------------------
\end{document}